\documentclass[aps,prl,twocolumn,preprintnumbers,groupedaddress,superscriptaddress,floatfix,tightenlines,reprint]{revtex4-1}
\usepackage{mathrsfs,natbib}
\usepackage{graphics,epsfig,color,subfigure}
\usepackage{verbatim}

\usepackage{graphicx,epsf,amssymb,bbm,amsbsy,amsfonts,amssymb,amsmath}
\usepackage{hyperref}

\newcommand{\eq}{\begin{equation}}
\newcommand{\eqe}{\end{equation}}

\newcommand{\eqa}{\begin{eqnarray}}
\newcommand{\eqae}{\end{eqnarray}}

\hypersetup{
    colorlinks=true,       
    linkcolor=red,          
    citecolor=blue,        
    filecolor=magenta,      
    urlcolor=blue           
}

\begin{document}

\title{$T$-Reflection}

\preprint{PUPT-2467, IPMU-14-0149}
\preprint{FTPI-MINN-14/14, UMN-TH-3337/14}

\author{G\"ok\c ce Ba\c sar}
\email{basar@tonic.physics.sunysb.edu}
\affiliation{Department of Physics and Astronomy,
Stony Brook University,
Stony Brook, New York 11794, USA}

\author{Aleksey Cherman}
\email{acherman@umn.edu}
\affiliation{Fine Theoretical Physics Institute, Department of Physics, University of Minnesota, Minnesota, MN 55455, USA}

\author{David A.\ McGady}
\email{dmcgady@princeton.edu}
\affiliation{Department of Physics, Jadwin Hall Princeton University Princeton, NJ 08544, USA.}

\author{Masahito Yamazaki}
\email{masahito.yamazaki@ipmu.jp}
\affiliation{Institute for Advanced Study, School of Natural Sciences, Princeton NJ 08540, USA}
\affiliation{Kavli IPMU (WPI), University of Tokyo, Kashiwa, Chiba 277-8586, Japan}
 
\begin{abstract}
We point out the presence of a $T \to -T$ temperature-reflection ($T$-reflection) symmetry for the partition functions of many physical systems. Without knowledge of the origin of the symmetry, we have only been able to test the existence of $T$-reflection symmetry in systems with exactly-calculable partition functions.  We show that $T$-reflection symmetry is present in a variety of conformal and non-conformal field theories and statistical mechanics models with known partition functions.  For example, all minimal model partition functions in two-dimensional conformal field theories are invariant under $T$-reflections. An interesting property of the $T$-reflection symmetry is that it can be broken by shifts of the vacuum energy. 
\end{abstract}


\maketitle

{\bf Introduction}---In this Letter we observe that the partition functions $Z(\beta = 1/T)$ of a number of physical systems have a curious symmetry under a `reflection' of the temperature, $T\to -T$:
\begin{align}
Z(+\beta) = e^{i \gamma} Z(-\beta) \, , 
\label{eq:Reflection}
\end{align}
where $e^{i \gamma}$ is a theory-dependent but temperature-independent complex number of modulus one. Note that asking about the behavior of $Z(\beta)$ under $T$-reflection requires information about all of the energy levels, $E_n$, and degeneracies, $d_n$, in
\eq
Z(\beta) = \sum_{n=1}^{\infty} d_n e^{-\beta E_n} \, .\label{Z_beta} 
\eqe
Naively, the sum representation of $Z(\beta)$ in Eq.~(\ref{Z_beta}) converges only for $\beta > 0$, and directly testing \eqref{eq:Reflection} by sending $\beta \to -\beta$ in $Z(\beta)$ hinges on being able to resum the series into some sort of closed form. Systems where such resummations are known are rare. 

In this Letter we consider a variety of quantum systems where exact expressions for $Z(\beta)$ are available and show that they have the $T$-reflection symmetry summarized in \eqref{eq:Reflection}. Many, but not all, of our example systems are conformal field theories (CFTs). The presence of $T$-reflection symmetry requires a unique choice of ground-state energy; shifting it by $-\Delta$ multiplies $Z(\beta)$ by a non-invariant factor of $e^{\beta \Delta}$. As $T$-reflection fixes the ground-state energy, it resembles a space-time symmetry. However, it seems to persist in interacting theories without conformal- or super-symmetry, and may be of help in resolving the cosmological constant problem~\footnote{Some potentially related ideas have appeared in e.g. \cite{Kaplan:2005rr,*tHooft:2006rs}}. 

Our most most intricate and surprising examples are two-dimensional CFT minimal models. Here, the role of the temperature is played by the modular parameter $\tau$ (${\rm Im}(\tau)>0$) of the torus, and $T$-reflection then acts on $\tau$ as \footnote{The same transformation was considered in, e.g., Ref.~\cite{Berkovich:1995nx}, albeit in a different context.}
\eq
R: \tau\to -\tau \, .
\eqe
We show that all characters of the Virasoro algebra which appear in any given minimal model are invariant under $R$-transformations/$T$-reflections. These characters, denoted $\chi_{r,s}^{p,p'}(q)$ and $\bar{\chi}_{m,n}^{p,p'}(\bar{q})$, are the building blocks of two-dimensional CFT minimal model torus partition functions, and hence all minimal models are invariant under $T$-reflections.  This $R$-transformation should be contrasted with the familiar modular transformation \cite{Cardy:1986ie} of the torus. The modular group $PSL(2, \mathbb{Z})$ is generated by two generators $S$ and $T$, which act on $\tau$ as
\eq
S:\tau\to -\frac{1}{\tau} \, , \quad T: \tau\to \tau+1 \, .
\eqe
The transformation $R$ does not commute with the generators of $PSL(2, \mathbb{Z})$. In fact $R,S,T$ generate the extended modular group  $PGL(2, \mathbb{Z})$.

{\bf Oscillators in Quantum Mechanics}---Let us begin with the simplest example, a bosonic simple harmonic oscillator. If we measure energy in units of the oscillator frequency, and denote the ground state energy by $\Delta$, the partition function is given by
\eq
Z_{\rm SHO}(\beta) = \sum_{n = 0}^{\infty} e^{-\beta(n+\Delta)}
= \frac{e^{\beta (1/2-\Delta)}}{2 \sinh (\beta/2)} \, .
\label{SHO} 
\eqe
This is $T$-reflection symmetric\label{Reflection}, with $\gamma=\pi$, only if the ground-state energy takes the naive value, $\Delta = +1/2$.

Similarly, the thermal partition function for a fermionic oscillator can be written as,
\eq
Z(\beta) =2 e^{\beta(1/2+\Delta)} \cosh(\beta/2) \, .
\label{fSHO}
\eqe
If $\Delta = -1/2$ this has a $T$-reflection symmetry ($\gamma = 0$). As emphasized above, shifting the ground-state energy breaks any symmetry under $T$-reflections.  One can also check that the twisted partition function $\tilde{Z} = \mathrm{Tr}\, (-1)^{\mathcal{F}} e^{-\beta H}$ (where $\mathcal{F}$ is fermion number) for the fermionic oscillator has $T$-reflection symmetry if and only if $\Delta=-1/2$.

{\bf Free $d=2$ CFTs}---Next we consider some of the simplest QFT examples of systems with $T$-reflection symmetry, which are free $d=2$ CFTs.  The torus partition functions of $d=2$ CFTs decompose into linear combinations of products of holomorphic and anti-holomorphic Virasoro characters, which can themselves be thought of as partition functions for left-moving and right-moving modes. Because $\tau = i \beta/L$ is the shape modulus of the torus --- $T^2 = S^1_L \times S^1_{\beta}$ --- and $q = e^{2\pi i \tau}$, the $T$-reflection properties of the partition functions are fixed by the transformation properties of the characters under $q \to q^{-1}$. 

We now examine the $q$-inversion properties of the single holomorphic character 
\eq
\chi_{\rm s}(q) 
 = q^{-\frac{1}{24}} \prod_{n = 1}^{\infty}\frac{1}{(1-q^n)} = \frac{1}{\eta(q)}\, . \label{2dBoson}
\eqe
contributing to the free scalar CFT (the analysis for anti-holomorphic sector is completely analogous).  We find
\eqa \label{2dBosonInv}
\chi_{\rm s}\left(q^{-1}\right) &=& q^{+\frac{1}{24}} \prod_{n = 1}^{\infty}\frac{1}{1-q^{-n}} \\
&=& \left\{ q^{\frac{1}{12}}
\prod_{a = 1}^{\infty} q^{a}
\prod_{b = 1}^{\infty} \frac{1}{(-1) } \right\} \chi_{\rm s}(q) \nonumber \\ 
&=&\frac{q^{\frac{1}{12}}
q^{\zeta(-1)}}{ (-1)^{\zeta(0)} } \chi_{\rm s}(q) = i \chi_{\rm s}(q) \, , \nonumber 
\eqae
where we used zeta-function regularization [$\zeta(s)$ is the Riemann zeta function].  So $\chi_{\rm s}(q)$ has $T$-reflection symmetry with $\gamma = \pi/2$.  

For additional insight into this result --- especially on the reason for the appearance of a regulator above --- we can compute the same partition function directly from the Euclidean path integral for a scalar field.  This gives
\eq
- \ln Z(\beta) 
=V_0 L \beta +\sum_{m=1}^{\infty}\left[  \frac{\beta \omega_m}{2}  + \log(1 - e^{-\beta \omega_m}) \right]
\label{eq:PathIntegralZ}
\eqe
where $\omega_m = 2\pi m/L$ and $V_0$ is the bare vacuum energy from the Lagrangian.  One can evaluate \eqref{eq:PathIntegralZ} as written, which yields $Z(\beta) = \chi_s(\beta)$, or evaluate \eqref{eq:PathIntegralZ} \emph{after} sending $\beta \to -\beta$, which yields $Z(-\beta) = \chi_s(-\beta)$.  Note that this amounts to what we did in the SHO example in \eqref{SHO} for each of the infinite number of oscillators in the field theory.  It is important to note that UV divergences appear in the calculations of both $Z(\beta)$ and $Z(-\beta)$ in QFTs, and hence the frequency sums must be regularized and renormalized, with the divergences absorbed in counterterms in {\it e.g.} $V_0$.   Indeed,  the second sum in \eqref{eq:PathIntegralZ} is precisely the same quantity as the infinite product in \eqref{2dBoson}, while the UV divergences and renormalization issues which appear in the first sum in \eqref{eq:PathIntegralZ} are hidden in the definition of the Casimir energy $\Delta = -1/24$ in \eqref{2dBoson}.  One can then verify that $Z(\beta)$ has $T$-reflection symmetry so long as the renormalized vacuum energy is set to the Casimir energy, just as in \eqref{2dBosonInv}, and also reproduce the $\gamma = \pi/2$ factor which we found using $\zeta$-function regularization above.   Similar remarks apply to all of our QFT examples.

We next check the free fermion CFT partition functions on $T^2$ (see {\it e.g.}~\cite{DiFrancesco:1997nk}).  We can take periodic (R) or anti-periodic (NS) boundary conditions for each of the two cycles of $T^2$, which yields three non-trivial distinct partition functions R-NS, NS-R, and NS-NS.  (R-R fermions have a zero mode which nullifies their associated partition function.) The difference between R-NS and NS-R comes because we take the Hamiltonian to be associated with isometries along one of the $S^1$'s in $T^2$. 

For NS-NS fermions, the character takes the form
\begin{align}
\chi_{\mathrm{NS-NS}}(q) = q^{-\frac{1}{48}} \prod_{n=0}^{\infty} (1+q^{n+1/2}) \, . \label{NSNS}
\end{align}
Under $T$-reflection, we have
\eqa
\chi_{\mathrm{NS-NS}}(q^{-1}) &=& q^{+\frac{1}{48}} \prod_{n=0}^{\infty} (1+q^{-(n+1/2)})  \\
&=&\left\{ q^{+\frac{1}{24}} \prod_{a=0}^{\infty} {q^{-(n+1/2)}} \right\} \chi_{\mathrm{NS-NS}}(q) \nonumber \\
&=&q^{\frac{1}{24}-\zeta(-1, \frac{1}{2})} \chi_{\mathrm{NS-NS}}(q) = \chi_{\mathrm{NS-NS}}(q) \, , \nonumber
\eqae
where $\zeta(s,a)$ is the Hurwitz zeta function, and $\zeta(-1,x) = -\frac{x^2}{2}+\frac{x}{2} -\frac{1}{12}$. 

Similar arguments allow us to verify that the R-NS and NS-R partition functions,
\begin{align}
\chi_{\rm{R-NS}}(q) &= q^{\frac{1}{24}} \prod_{n=0}^{\infty} (1+q^n) \, , \label{RNS} \\
\chi_{\mathrm{NS-R}}(q) &= q^{-\frac{1}{48}} \prod_{n=1}^{\infty} (1-q^{n-1/2}) \, , \label{NSR}
\end{align}
are each separately covariant under $T$-reflection.

{\bf Free gauge theories in $d=4$-dimensions}---We now consider gauge theories on $S_R^3 \times S^1_{\beta}$ where $R$ and $\beta$ are the radii of the 3-sphere and the thermal circle respectively.  Asymptotically-free gauge theories with gauge groups of rank $N$ with a strong scale $\Lambda$ become arbitrarily weakly coupled when $R \Lambda \to 0$.  Indeed, in the $R\Lambda \to 0$ limit they behave as if they were free CFTs on $S^3_{R} \times S^1_{\beta}$ with a color-singlet constraint from the Gauss law.  Refs.~\cite{Sundborg:1999ue,*Polyakov:2001af,*Aharony:2003sx} showed how to compute their partition functions in this free CFT limit.  The gauge theory partition functions can be written in terms of so-called `single-particle' partition functions, which can be calculated using CFT techniques and are given by
\eqa
&& z_{S}(x)= \frac{(x^{1/2}+x^{-1/2})}{(x^{-1/2}-x^{1/2})^{3}} \, ,\label{eq:letterS} \\
&& z_{F}(x) =\frac{2^3}{(x^{-1/2}-x^{1/2})^3} \, ,\label{eq:letterF} \\
&& z_{V}(x) = 1+\frac{(x^{2}-x^{-2})-4(x-x^{-1})}{(x^{-1/2}-x^{1/2})^4} \, \label{eq:letterV}
\eqae
for real scalars, Majorana fermions, and vectors respectively and $x = e^{-\beta/R}$.   We observe that $z_S, z_F$ and $1-z_V$ are $T$-reflection symmetric with  $\gamma =\pi$.\footnote{Note that while the free Maxwell theory is conformal only in $d = 4$ \cite{ElShowk:2011gz}, but $z_S$ and $ z_F$ can be computed in any $d>2$ via the state-operator correspondence, and they transform with $\gamma=(d-1)\pi$ under $T$-reflection. }

The single-trace and full multi-trace confining-phase partition functions of (nearly-free) large-$N$ gauge theories with adjoint matter on $S_R^3 \times S_\beta^1$ have $T$-reflection symmetry, as they depend only on the covariant functions $1-z_V(x)$, $z_S(x)$, and $z_F(x)$.\cite{Sundborg:1999ue,*Polyakov:2001af,*Aharony:2003sx} For example, for a theory with $n_f$ and $n_s$ massless adjoint fermions and conformally-coupled scalars, the confining-phase thermal partition function of the gauge theory in the large-$N$ limit can be written as 
\eq
Z_{G}(\beta) = \prod_{n=1}^{\infty}{1\over 1 - z_V(x^n)- n_s z_S(x^n) + (-1)^n n_f z_F(x^n)} \label{fullZ}.
\eqe
The values $n_s=6$ and $n_F=4$ correspond to $\mathcal N=4$ Super Yang-Mills theory. $T$-reflection maps each single particle partition function, \eqref{eq:letterS}-\eqref{eq:letterV}, and hence each factor in the product \eqref{fullZ}, into itself, up to a factor of $(-1)$.  To compute the value of $\gamma$ for the transformation of $Z_{G}$ we must deal with the formal expression
\eq
\prod_{n=1}^{\infty} (-1) = (-1)^{\sum_{n=1}^{\infty} 1} \equiv (-1)^{\zeta(0)} = e^{-i \pi/2}
\eqe
where in the last step we used zeta function regularization to define the value of the infinite sum.  Hence these gauge theories are $T$-reflection symmetric with $\gamma = \pi/2$.

We have checked that the free conformally-coupled massless scalar, massless fermion, and $U(1)$ gauge theories without matter on $S^3 \times S^1$ all have $T$-reflection symmetry, but only if the vacuum energy is set to the appropriate Casimir energy.  Since $Z_G$ describes a family of confining theories\cite{Aharony:2003sx}, one might have expected that the vacuum energy would be of order the confinement scale $1/R$, which in this case is also the order of a Casimir energy on $S^3$.  However,  from \eqref{fullZ} one can see that $Z_{G}$ has $T$-reflection symmetry only if its vacuum energy \emph{vanishes}.  We do not know the reason for this phenomenologically tantalizing result.

{\bf Superconformal Indices}---Next we consider twisted partition functions on $\mathcal{M} \times S^1_{\beta}$ for supersymmetric field theories, where the compact manifold $\mathcal{M}$ and the boundary conditions on the circle $S^1_{\beta}$ are chosen such that supersymmetry is preserved.  This requires that fermions have periodic boundary conditions on $S^1_{\beta}$, and $S^1_{\beta}$ can be interpreted as a spatial circle. Our discussion applies to  $d = 4, \, \mathcal{N}= 1$ theories with arbitrary matter content (with the rank of the gauge group $N$ kept finite), as long as the theory has a $U(1)$ R-symmetry.

The virtue of such generalized partition functions is that they are independent of the continuous coupling constants and can be evaluated exactly even for strongly interacting theories, by taking the free coupling limit. We again essentially have a set of decoupled oscillators. In the examples we have been able to check, these partition functions are also invariant under $T$-reflections.

For $\mathcal{M}=S^3$, we have the so-called superconformal index \cite{Kinney:2005ej,*Romelsberger:2005eg} defined for \ $d=4$, $\mathcal{N}= 1$ gauge theories on $S^3 \times S^1$,  which is defined to be 
\begin{align}
I=\textrm{Tr}\left[(-1)^{\mathcal{F}} p^{\frac{E+j_2}{3}+j_1} q^{\frac{E+j_2}{3}-j_1} \prod_i u_i^{F_i} \right] \, ,
\end{align}
where $E, j_1, j_2$ are the Cartan generators of $U(1)\times SU(2)\times SU(2)$ isometry of $S^3\times S^1$, $F_i$ are the flavor Cartan generators of the theory, and $ p, q, u_i$ are the relevant fugacities. These fugacities depend exponentially on $\beta$, and $T$-reflection inverts them. 

As tabulated in e.g.~\cite{Kinney:2005ej,Dolan:2008qi}, the single-particle indices for vector-multiplets $z_V$ and chiral-multiplets $z_S$ are
\eqa\label{SCI}
&&z_V(p,q) = 1 - \frac{(pq)^{1/2} + (pq)^{-1/2}}{(p^{1/2}-p^{-1/2})(q^{1/2}-q^{-1/2})} \, ,\, \nonumber\\
&&z_S(p,q,u) = \frac{(pq)^{1/2}u^{-1} + (pq)^{-1/2} u}{(p^{1/2}-p^{-1/2})(q^{1/2}-q^{-1/2})} \, .
\eqae
Although $T$-reflections invert the fugacities, we see that the rational functions $(1-z_V)$ and $z_S$ are invariant.

As was the case for the free CFTs on $S_R^3 \times S^1_{\beta}$ above, the full superconformal indices are functions of $z_S$ and $1-z_V$. From this one can show that the entire superconformal index, a quantity associated with \emph{interacting} SUSY gauge theories, has a $T$-reflection symmetry with $\gamma = 0$. 

We have also verified that a similar argument works for the more general indices on $S^1 \times S^3/\mathbb{Z}_p$ from ~\cite{Benini:2011nc}, which in turn implies a $T$-reflection symmetry for their dimensionally-reduced counterparts (e.g.\ three-dimensional index on $S^1\times S^2$ \cite{Imamura:2011su}).   For $S^1 \times S^3/\mathbb{Z}_p$, $T$-reflection also flips the discrete holonomies along $S^3/\mathbb{Z}_p$.

{\bf  Minimal models}---We next consider $d=2$ minimal model CFTs, which give an infinite family of interacting but solvable field theories.  These CFTs describe the critical behavior of a wide variety of models from statistical physics, many of which are experimentally realizable. A minimal model $\mathcal{M}(p,p')$ has a finite set of primary operators, characterized by two integers $r$ and $s$, satisfying $1\le r < p, 1\le s<p'$. The characters of minimal models are given by \cite{Rocha-Caridi,*Feigin:1982tg}
\eqa
&&\chi_{(r,s)}^{p,p'}= K^{(p,p')}_{r,s}(q) - K^{(p,p')}_{r,-s}(q) \, , \label{Character} \\
&&K^{(p,p')}_{r,\pm s}(q) = \frac{1}{\eta(q)} \sum_{n = -\infty}^{+\infty} q^{\frac{(2pp' \, n \, + (pr \mp p's))^2}{4pp'}} \, .  \label{Karacter} 
\eqae
The sum in \eqref{Karacter} takes the form of a theta function---$\vartheta_{00}(z,Q)$---which has an infinite product representation:
\begin{eqnarray}
&& \eta(q)K^{(p,p')}_{r,\pm s}(q) = q^{\frac{(pr \mp p's))^2}{4pp'}} \sum_{n = -\infty}^{+\infty} (q^{2pp'}) ^{\frac{n^2}{2}} (q^{pr \mp p's}) ^{n} \nonumber\\
&& \,\,\,\, = Q^{\frac{\alpha^2}{2}} \prod_{n = 1}^{\infty} (1-Q^n) (1+ Q^{n+\alpha-\frac{1}{2}}) (1+ Q^{n-\alpha-\frac{1}{2}}) \label{Theta2}
\end{eqnarray}
where $Q = q^{2 pp'}$, $\alpha = \frac{pr \mp p's}{2pp'}$, and we used Jacobi's triple-product identity for $\vartheta_{00}(Q^{\alpha},Q)$ in the last line. $Q(q)$-inversion maps $\eta(q)K^{(p,p')}_{r,\pm s}(q)$, i.e. $ Q^{\frac{\alpha^2}{2}} \vartheta_{00}(Q^{\alpha}, Q)$, into
\eqa
\label{theta_inversion}
Q^{-\frac{\alpha^2}{2}}\vartheta_{00}(Q^{-\alpha},Q^{-1}) &=& \frac{Q^{-\frac{\alpha^2}{2}} \vartheta_{00}(Q^{+\alpha},Q^{+1}) (-1)^{\zeta(0)}}
{Q^{\zeta(-1)+\zeta(-1, \frac{1}{2}-\alpha)+\zeta(-1, \frac{1}{2}+\alpha)}}\nonumber \\
&=& -i \, Q^{+\frac{\alpha^2}{2}}\vartheta_{00}(Q^{\alpha},Q) \, .
\eqae
The $T$-reflection symmetry of the character $\chi^{p,p'}_{(r,s)}$ then follows immediately from Eqs.~\eqref{2dBoson}, \eqref{2dBosonInv}, \eqref{Theta2} and \eqref{theta_inversion}, and we find $\gamma=0$. We have carried out similar checks for the characters of $\mathcal{N}=1, 2$ super-minimal models. All these characters have $T$-reflection symmetry.   

These examples can be run backwards to show that $T$-reflection invariance fixes the ground state energy to take the values which coincide with the values mandated by other physical principles, such as modular invariance. 

{\bf Lattice Models}---At least some exactly solved lattice models in two-dimensional statistical mechanics also have $T$-reflection symmetry. For instance, Onsager's exact solution \cite{Onsager} of the two-dimensional Ising model on the square lattice, with periodic boundary conditions and zero external field, is
\eqa
{\rm ln}\left[Z(\beta)\right] &=& N\, {\rm ln}(2 {\rm cosh} 2 \beta J) \nonumber\\ 
&+& \frac{N}{\pi} \int_{0}^{\pi/2}\!
\! dw \,\, {\rm ln} \left[ \frac{1}{2} \left\{ 1 + (1- K^2 {\rm sin}^2 w)^{1/2} \right\} \right]\, , \nonumber\\
K &=& \frac{2 {\rm sinh} (2 \beta J)}{({\rm cosh} (2 \beta J))^2} \, ,
\label{eq:Ising}
\eqae
where $J$ is the nearest neighbor interaction, and $N$ is the number of sites on the lattice ($N\gg 1$). This solution has the $T$-reflection symmetry for arbitrary $T$, even away from the critical point. (At the critical point the Ising model is described by the minimal model $\mathcal{M}(4,3)$, for which we already verified $T$-reflection.)  Note that this is an explicit example of an interacting non-supersymmetric and non-conformal many-body model with $T$-reflection symmetry.  The symmetry would be broken if there is a shift of the ground state energy in \eqref{eq:Ising}.  

As shown in Refs.~\cite{Yamazaki:2012cp,*Terashima:2012cx,*Yamazaki:2013nra},  superconformal indices of $\mathcal{N}=1$ quiver gauge theories can be identified with partition functions of two-dimensional exactly solvable statistical mechanics models. Exploiting the $T$-reflection symmetry of the $\mathcal{N} = 1$ four-dimensional superconformal indices explored above, one could show a wide variety of models in two-dimensional statistical mechanics also have $T$-reflection symmetry.

{\bf Discussion}---Several comments are now in order.

It is important to keep in mind that $T$-reflection symmetry can only hold for a special class of theories; for arbitary $E_n$ the partition function \eqref{Z_beta} cannot have any simple $T$-reflection transformation. For instance, a three level system with unevenly spaced energy levels will not be $T$-reflection symmetric.  For interacting systems with an infinite number of energy levels it is not easy to check $T$-reflection, as we have already remarked in the introduction.  One of the exceptions is the case where $E_n$ is a quadratic polynomial in $n$, where we can we can relate the transformation of \eqref{Z_beta} to the inversion properties of theta functions.  While many of our examples were free theories, we again emphasize that $T$-reflection symmetry is not necessarily spoiled by interactions, as is highlighted by the minimal model examples.

It is conceivable that in general $T$-reflection symmetry may require simultaneous transformations of other parameters of the system. As a simple example, consider adding an anharmonic interaction $\frac{\lambda}{4!} \, \phi^4$ to the harmonic oscillator. The first-order correction to the partition function is (e.g.\ Ref.~\cite{Naya})
\begin{align}
Z(\beta, \lambda)  = Z_{\rm SHO}(\beta)\left[1-\frac{\beta \lambda(1+x)^2}{8(1-x)^2} \right] \ ,
\end{align}
where $x = e^{-\beta}$ and $Z_{\rm SHO}$ was given earlier in \eqref{SHO} (we take $\Delta = 1/2$).  One then finds that $Z\to -Z$ under $\beta \to -\beta, \lambda \to -\lambda$.  Note that the expression above has the $\lambda$-corrected vacuum energy
\begin{align}
\Delta(\lambda) = \frac{1}{2}\left(1 + \frac{\lambda}{4}\right) .
\end{align}
Just as in the non-interacting theory, one is not allowed to shift the vacuum energy from this particular value without ruining the $T$-reflection `symmetry'. Similar conclusions seem to hold in higher orders of perturbation theory.   Nonetheless, it is uncertain whether conclusions based on finite-order perturbation theory carry any real information in this context: $T$-reflection covariance depends crucially on the structure of the entire spectrum. Higher-lying states in the spectrum depend more sensitively on the anharmonicities. So it is an open question whether perturbative studies of deformations of exactly-solvable systems will have definitive lessons for the survival of $T$-reflection symmetry in more general settings.

All of our examples and remarks raise several obvious interrelated questions: what is the origin of the $T$-reflection symmetry? How general is the symmetry? What are the broader implications of $T$-reflection symmetry? The answers to these questions are currently unknown.    It is conceivable that $T$-reflection symmetry is a previously unappreciated consequence of known symmetries, such as e.g. time-reversal symmetry, or it might be something entirely new.   All we can say for now is that $T$-reflection symmetry is not a consequence of conformal invariance, nor of supersymmetry, since we have given examples of $T$-reflection symmetric systems which are neither conformal nor supersymmetric.

In some examples the reflection symmetry has a simple group-theoretical explanation. As an example, consider the character of the spin $j$-representation of $SU(2)$:
\begin{align}
\textrm{Tr}\, q^{J_3} = \frac{q^{2j+1} -q^{-2j-1}}{q-q^{-1}} \, ,
\end{align}
This is invariant under $q\to q^{-1}$, which amounts to an exchange of the highest-weight state with the lowest-weight state.  So here $T$-reflection is an involution (inner automorphism) of the $SU(2)$ algebra which exchanges the raising and lowering operators $J_{\pm}:=J_1\pm i J_2$ via $J_3\to -J_3 \, , \quad J_{\pm} \to J_{\mp}$, which is simply a $\pi$ rotation along the axis $1$. Hence, for $SU(2)$, $T$-reflection symmetry is implied directly by the the algebraic structure itself. Similar involutions (called Cartan-Chevalley involutions) exist for arbitrary semisimple Lie algebras. One might then naturally wonder whether $T$-reflection symmetry for CFTs might have something to do with the fact that the Virasoro algebra
\begin{align}
[L_m, L_n]=(m-n)L_{m+n} +\frac{c}{12}\delta_{m+n,0} \, ,
\end{align}
has an outer automorphism $L_n\to -L_{-n}$. Such group-theory-based arguments do not seem to have a chance to explain $T$-reflection in general, however. The involution exchanges the highest energy state with the lowest energy state, but many of our examples correspond to infinite-dimensional representations without highest energy states.

We would like to thank A. Altu\u g, N.\ Arkani-Hamed, J.\ L.\ Evans, Y.-t.\ Huang, Z.\ Komargodski, J.\ Maldacena, Y.\ Oshima, A.\ Polyakov, H.\ Verlinde, A.\,Zhiboedov for discussions. This work is supported by the U.S. Department of Energy under the grants  DE-FG-88ER40388 (G.\ B.), DE-SC0011842 (A.\ C.).

\bibliographystyle{apsrev4-1}
\bibliography{Treflection.bib} 

\end{document}